\def\rfr#1{(\ref{#1})}

\def\Rfr#1{(\ref{#1})}

\def\bar{\begin{eqnarray}}
\def\ear{\end{eqnarray}}

\def\eqi{\begin{equation}}
\def\eqf{\end{equation}}
\def\eqia{\begin{eqnarray}}
\def\eqfa{\end{eqnarray}}
\def\rp#1#2{{#1\over#2}}

\def\lb#1{\label{#1}}

\documentclass[11pt]{article}
\usepackage{amsmath,amsthm,amscd,amssymb}
\usepackage{latexsym}
\usepackage{graphicx,epsfig}

\begin{document}

\noindent{\bf \LARGE{Are We Far from Testing General Relativity
with the Transiting Extrasolar Planet HD 209458b `Osiris'?}}
\\
\\
\\
{L. Iorio}\\
{\it Viale Unit$\grave{a}$ di Italia 68, 70125\\Bari, Italy
\\tel./fax 0039 080 5443144
\\e-mail: lorenzo.iorio@libero.it}

\begin{abstract}
In this paper we investigate the possibility of measuring the
general relativistic gravitoelectric correction $P^{\rm (GE)}$ to
the orbital period $P$ of the transiting exoplanet HD 209458b
'Osiris'. It turns out that the predicted magnitude of such an
effect is $\sim 0.1$ s, while the most recent determinations of
the orbital period of HD 209458b with the photometric transit
method are accurate to $\sim 0.01$ s. It turns out that the major
limiting factor is the error in the measurement of the projected
semiamplitude of the star's radial velocity $K_M$. Indeed, it
affects the determination of the mass $m$ of the planet which, in
turn, induces a systematic error in the Keplerian period $P^{(0)}$
of $\sim 8$ s.
However, the present situation may change if future one-two orders
of magnitude improvements in the Doppler spectroscopy technique
will occur.
\end{abstract}

\section{Introduction}
In this paper we investigate the possibility of performing a test
of general relativity, in its weak-field and slow-motion
approximation, for the first time in a planetary system other than
our Solar System\footnote{See e.g.
\texttt{http://www.obspm.fr/encycl/} for regularly updated
information about the extrasolar planets.}.

More precisely, it has been shown (Soffel 1989; Mashhoon et al.
2001) that the general relativistic gravitoelectric (GE) part of
the gravitational field of a static, spherically symmetric
mass\footnote{It is the Schwarzschild component of the space-time
metric tensor which gives rise to various tested phenomena (Will
1993) like the well known secular Einstein pericenter advance, the
deflection of the electromagnetic waves, the Shapiro time delay
and the de Sitter geodetic precession of the spin of an orbiting
gyroscope.} also induces an additional small correction $P^{\rm
(GE)}$ to the Keplerian orbital period $P^{(0)}$ of a test
particle. For a circular orbit of radius $a$ around a body of mass
$M$ it is, to order $\mathcal{O}(c^{-2})$ \eqi P^{\rm
(GE)}\sim\rp{3\pi}{c^2}\sqrt{GMa},\lb{mash}\eqf where $G$ is the
Newtonian gravitational constant and $c$ is the speed of light in
vacuum. Such an effect has not yet been directly measured. The
possibility of detecting it in the Sun-Mercury system, for which
it amounts to $P^{\rm (GE)}\sim 0.3 $ s, has preliminarily been
studied in (Iorio 2005).

At first sight rather surprisingly, it might be easier to detect
$P^{\rm (GE)}$ in some of the recently discovered exoplanets than
in our Solar System. Indeed, many of the exoplanets whose
parameters and ephemerides are best known have masses comparable
to that of Jupiter and orbital periods of just a few days. This
allows for a high accuracy--to a subsecond level in some cases--in
measuring their periods from combined photometric transit and
spectroscopic radial velocity measurements. Indeed, it is more
likely to observe a transit if the orbital period is short.
Moreover, the magnitude of the disturbing effects can only be
observed if the main star is not too massive and the planet is not
too light. Of course, such kind of measurements could not be
performed in the case of Mercury due to its extremely tiny
perturbing effect on the Sun and its relatively long period of
almost 88 days. Only difficult and rather sparse astrometric
measurements could be done.

The possibility of testing other relativistic signals which affect
the orbital motion, like the gravitoelectric  secular precessions
of the periastron $\omega$ and of the mean anomaly $\mathcal{M}$,
and the gravitomagentic secular precessions of $\omega$ and of the
longitude of the ascending node $\Omega$, must be ruled out
because the ephemerides of the extrasolar planets do not currently
allow to determine all the details of the orbital paths except for
the orbital period.
\section{The Two-Body Relativistic Correction to the Orbital
Period}\lb{kazonga} In the first post-Newtonian approximation, the
motion of two particles of rest masses $M$ and $m$, each one
moving in the gravitational field of the other, can be studied in
the post-Newtonian center of mass frame (Soffel 1989). In the
standard isotropic coordinate\footnote{It is the coordinate used
in the force models of the post-Newtonian equations of motion
(Estabrook 1971) adopted, e.g., for the computation of the
planetary ephemerides by the Jet Propulsion Laboratory (JPL). }
$r$, not to be confused with the Schwarzschild radial coordinate
$r^{'}$ which, for $M>>m$, is $r^{'}=r(1+GM/2c^2 r)^2$, the
general relativistic  equation of motion is (Soffel 1989)
{\small{\eqi \ddot{\bf{r}}=-\rp{\mu}{r^3}{\bf{r}}+\rp{\mu}{c^2 r^3
}\left\{\left[\rp{\mu}{r}(4+2\nu)-(1+3\nu){\dot{\bf{r}}}^2+\rp{3\nu}{2r^2}({\bf{r}}\cdot\dot{\bf{r}})^2\right]
{\bf{r}}+({\bf{r}}\cdot\dot{\bf{r}})(4-2\nu)\dot{\bf{r}}\right\},\lb{accel}\eqf}}
where $\mu\equiv G(M+m)$, $\nu\equiv Mm/(M+m)^2$ and the overdot
represents the time derivative.

In the case of circular orbits, i.e.
${\bf{r}}\cdot\dot{\bf{r}}=0$, which often occurs for many
exoplanets orbiting close to their star due to tidal
circularization, the post-Newtonian acceleration is entirely
radial. In polar coordinates \rfr{accel} reduces to
\eqi r\dot\phi^2=\rp{\mu}{r^2}-\rp{\mu}{c^2 r^2
}\left[\rp{\mu}{r}(4+2\nu)-(1+3\nu)(r\dot\phi)^2\right].\lb{accel2}\eqf
The orbital period can easily be calculated from  \rfr{accel2} for
$r=$const$\equiv a$.  From \eqi \dot\phi=\sqrt{
\rp{\mu}{a^3}\left[\rp{ 1-\rp{\mu}{c^2 a }(4+2\nu) }{
1-\rp{\mu}{c^2 a}(1+3\nu) }\right] }\eqf one gets that the time
$P$ required for $\phi$ in describing an angular interval of
2$\pi$ with respect to a fixed direction is $P=P^{\rm (0)}+P^{\rm
(GE) }$
with \begin{eqnarray} P^{(0)}&=&2\pi\sqrt{\rp{a^3}{\mu}},\lb{kep}\\
P^{\rm (GE)}&=&\rp{3\pi}{c^2}\sqrt{\mu
a}\left(1-\rp{\nu}{3}\right)+\mathcal{O}(c^{-4})\lb{twobody}.
\end{eqnarray}
Note that for $M>>m$, i.e. $\mu\rightarrow M$ and $\nu\rightarrow
0$, \rfr{twobody} reduces to \rfr{mash} (Mashhoon et al 2001).
Moreover, \rfr{twobody} agrees with the expression (A2.50)
obtained by Soffel (1989) in the Brumberg representation.

\section{What Kind of Exoplanets are Suitable for Better
Measuring the Relativistic Correction to the Orbital Period?} Let
us evaluate the orders of magnitude involved in a typical
exoplanetary scenario with $M={\rm M}_{\odot},$ $m={\rm m}_{\rm
Jup},$ $a=0.1$ astronomical units (AU); the term
$(3\pi/c^2)\sqrt{\mu a}\sim 0.1 $ s, while $\nu/3\sim 10^{-4}$, so
that it can safely be neglected in the subsequent analysis because
it would be undetectable.
%
In order to measure $P^{(\rm GE)}$ two requirements must be
fulfilled
\begin{itemize}
  \item The planet must be as closest as possible to its star,
  i.e. $a$ must be as smallest as possible. Indeed, in this case
  the orbital period is short and a large number of orbital
  revolutions can be recorded, thus increasing the observational
  accuracy. The observational error $\sigma_{P^{\rm (obs)}}$ must typically be smaller than
  $0.1$ s $= 1\times 10^{-6}$ days.
  \item The Keplerian period $P^{(\rm 0)}$ must be subtracted from the observed period in
  order to single out the relativistic component as $P^{(\rm GE)}=P^{\rm (obs)}-P^{(0)}$. As a consequence,
  the uncertainty in $P^{(0)}$ due to the errors in the measured
  planet's mass and semimajor axis induces a systematic bias
  which affects the total error
  $\sigma^{(\rm tot)}_{P^{(\rm GE)}}=\sigma_{P^{\rm (obs)}}+\sigma_{P^{(0)}}$. It turns out
  \eqia\rp{P^{(\rm
  GE)}}{\left.\sigma_{P^{(0)}}\right|_{\mu}}&=&\rp{3}{c^2}\rp{\mu^2}{a\sigma_{\mu}},\\
  \rp{P^{(\rm GE)}}
  {\left.\sigma_{P^{(0)}}\right|_a}&=&\rp{\mu}{c^2\sigma_a}.\eqfa
  Thus, in order to reduce the impact of
  $\sigma_{P^{(0)}}$ the mass of the system must be as largest as
  possible, the semimajor axis must be as smallest as possible
  and, of course, the errors in $a$ and $\mu$ must also be small.
\end{itemize}

In regard to the observational accuracy, the so far  discovered
exoplanets whose periods have been measured with a precision close
to our stringent requirements are TrES-1 ($\sigma_{P^{\rm
(obs)}}=8\times 10^{-6}$ days) (Alonso et al. 2004), OGLE-TR-56
($\sigma_{P^{\rm (obs)}}=5.9\times 10^{-6}$ days) (Konacki et al.
2003) and HD209458b ($\sigma_{P^{\rm (obs)}}=1.8\times 10^{-7}$
days) (Wittenmyer et al. 2005).
\section{The HD 209458b System}\lb{OSIRIS}
The most interesting system, for our purposes, is thus HD 209458b
'Osiris' which, of more than 150 recently discovered extrasolar
planets, is the first known to transit its star (Charbonneau et
al. 2000; Henry et al. 2000). The most recent determinations of
its parameters and ephemerides, based on twenty-seven transit
events in four bandpasses and on more than three years of
high-precision radial velocities, are due to Wittenmyer et al.
(2005).

Its main star, 47 pc far from us, has a mass which has been
estimated by Cody \& Sasselov (2002) to lie in the range $M/{\rm
M}_{\odot}=1.06\pm 0.13$. It is important to note such a range
does not come from a direct measurement, i.e. $M$ is not one of
the fifteen system's parameters simultaneously fitted in the
global data reduction of HD 209458 by means of the Eclipsing Light
Curve (ELC) code (Orosz \& Hauschildt 2000). It is an interval
based on stellar evolution models and measurements of luminosity
and temperature: no dynamical effects related in some ways to
relativistic orbital motions have been used.  $M$ has been held
fixed in (Wittenmyer et al. 2005) and different values for its
mass have been adopted.

The multi-color transit light curves allow to measure, among other
things, the inclination
$i$ which amounts to 86.67 deg (Wittenmyer et al. 2005).

The measurement of the radial velocity $K_M$ curve and of the
orbital period $P$ allows to determine the mass function $f$
from which the planet's mass $m$ can, thus, be deduced, for a
given value of the stellar mass $M$. The planetary mass ranges
from $0.590$ m$_{\rm Jup}$ ($M$/M$_{\odot}$=0.93) to $0.697$
m$_{\rm Jup}$ ($M$/M$_{\odot}$=1.19) (Wittenmyer et al. 2005).
Note that, up to now, the adopted model for $f(m)$ is entirely
Newtonian; see Section \ref{appe} for a modified version also
including relativistic corrections.

Finally, the semimajor axis $a$ can be obtained from the relation
which links it to the measured orbital period $P^{\rm (obs)}$ and
the system's masses. The Kepler's third law \rfr{kep} is used to
this aim. The so obtained values in (Wittenmyer et al. 2005) range
from 0.044 AU ($M/{\rm M}_{\odot}=0.93$) to 0.048 AU ($M/{\rm
M}_{\odot}=1.19$).

In regard to $P^{\rm (GE)}$, it turns out that, by using
\rfr{mash}, the adopted range for $M$ and the measured values for
$a$ it ranges from 0.095 s to 0.112 s. Note that the assumption of
a circular orbit for HD 209458b, adopted also by Wittenmyer et al.
(2005) in their analysis, is consistent with a tidal
circularization time of order $10^8$ years (Bodenheimer, Laughlin
\& Lin 2003) and radial velocity observations (Mazeh et al. 2000).

The orbital period is measured independently of any gravitational
theory model as one of the fifteen parameters of the fit to the
observations with the ELC software. Such important parameter has
been determined with high accuracy by combining the ground-based
observations, the transits observed with the photomultiplier tubes
(PMTs) in the Fine Guidance Sensor (FGS) of the Hubble Space
Telescope (HST) and the transits observed with the Space Telescope
Imaging Spectrograph (STIS) over many cycles.
The quoted value in (Wittenmyer et al. 2005) is \eqi P^{\rm
(obs)}_{\rm (Wittenmyer\ et\ al.) }=3.52474554\pm 1.8\times
10^{-7} {\rm days}.\eqf The uncertainty in $P$ is thus 0.016 s.
Similar results have also been obtained by Schultz et al. (2004)
who quoted \eqi P^{\rm (obs)}_{\rm (Schultz\ et\ al.)
}=3.52474408\pm 2.9\times 10^{-7} {\rm days},\eqf with an
uncertainty of 0.025 s. The situation is, thus, now more favorable
than that described in (Iorio 2005) in which the exoplanet
OGLE-TR-132b was examined.

However, it must be noted that there is still a discrepancy of
0.126 s between the two measurements: it is as large as the
relativistic effect itself. Such an uncertainty is due to the fact
that, despite their very high photometric precision, neither the
FGS or STIS can obtain an uninterrupted observation of the transit
due to the low orbit of the HST. The data from the Canadian
Microvariability and Oscillations of STars (MOST) satellite,
launched in June 2003, should overcome these problems enabling to
continuously observing HD 209458 for many transits with
exceptionally high precision (Rucinski et al. 2003).

The fact that the predicted relativistic component of the orbital
period $P^{\rm (GE)}$ falls into the measurability domain for HD
209458b is very important because it opens the possibility of
measuring it provided that the obtainable precision is high
enough. As a consequence, it may turn out that relativistic
corrections should be accounted for, e.g., in modelling both the
mass function and the orbital period in order to refine the
measurements of $m$ and $a$. In Section \ref{appe} we will deal
with such topics.
\section{A Modified Mass Function to Order $\mathcal{O}(c^{-2})$}\lb{appe}
In the barycentric frame the relation \eqi M a_M=m a_m\lb{app}\eqf
can be assumed for the star and its planet: indeed, the
post-Newtonian corrections of order $\mathcal{O}(c^{-2})$ to it
are negligible because are proportional to (Wex 1995)
\eqi\epsilon_{1\rm
PN}=\frac{1}{c^2}\frac{m}{M}\frac{1}{\left(1+\frac{m}{M}\right)^3}\left[v^2-\frac{G(m+M)}{r}\right],\lb{cazzo}\eqf
where $v$  and $r$ are the relative velocity and distance.
Moreover, for circular orbits, as in this case, \rfr{cazzo}
vanishes. By adding $m a_M$ to both members of \rfr{app} it is
possible to straightforwardly obtain \eqi a^3_M\sin^3 i= \rp{a^3
m^3\sin^3 i}{(M+m)^3},\lb{merdo}\eqf where $a=a_M+a_m$. By using
the relation \eqi K_M=\left(\rp{2\pi a_M}{P}\right)\sin
i=\left(\rp{m}{M+m}\right)\left(\rp{2\pi a}{P}\right)\sin
i,\lb{KM}\eqf where $K_M$ is the projected semiamplitude of the
star's radial velocity, is it possible to derive the general
relation \eqi\frac{K_M^3 P^3}{8\pi^3}=\frac{a^3m^3\sin^3
i}{(m+M)^3}.\lb{vgu}\eqf Note that the left-hand side of \rfr{vgu}
is a phenomenologically determined quantity.

If we use the third Kepler's law \eqi
P^2=4\pi^2\rp{a^3}{G(M+m)}\eqf to model the orbital period $P$ it
is possible to obtain the so called mass function $f(M, m)$ as
\eqi f\equiv \rp{PK^3_M}{2\pi G}=\rp{m^3\sin^3
i}{(M+m)^2},\lb{massf}\eqf For HD 209458b it amounts to $82.7\pm
1.3$ m s$^{-1}$ (Wittenmyer et al. 2005).
The relation \Rfr{massf} is of the utmost importance because it is
possible to measure the planet's mass $m$ by keeping $M$ fixed and
measuring $i, P$ and $K_M$ from the photometric transit curve and
Doppler spectroscopy.

The present-day observational accuracy in measuring the period $P$
of HD 209458b suggests that general relativistic corrections might
play a role in the system's parameters determination process.
Thus, we will now derive a modified expression of the right-hand
side of \rfr{massf} to order $\mathcal{O}(c^{-2})$. To this aim,
let us write
\eqi P\sim 2\pi\sqrt{\rp{a^3}{\mu}}+\rp{3\pi}{c^2}\sqrt{\mu
a}.\lb{picc}\eqf
By defining
  \eqi \left\{
\begin{array}{lll}
    x &\equiv& a,\\\\
    q &\equiv& \rp{4\pi^2}{\mu},\\ \\
    b &\equiv& \rp{12\pi^2}{c^2},\\\\
    d &\equiv& P^2,
  \end{array}
\right. \eqf \rfr{picc} can be written to order
$\mathcal{O}(c^{-2})$ as \eqi qx^3+bx^2=d.\lb{minchia}\eqf  Let us
rewrite \rfr{minchia} as \eqi qx+b=\rp{d}{x^2};\lb{frik}\eqf since
$q,b,d,x$ are positive it admits a solution which is also unique.
Let us look for a solution of the form \eqi
x=x^{(0)}(1+\epsilon),\lb{eppis}\eqf with $\epsilon\ll 1$ because
of order $\mathcal{O}(c^{-2})$ and \eqi
x^{(0)}=\left(\rp{d}{q}\right)^{1/3}. \eqf Inserting \rfr{eppis}
into \rfr{frik} yields, to order $\mathcal{O}(c^{-2})$
\eqi\epsilon\sim-\rp{b}{3x^{(0)}q}.\eqf Thus, \eqi x\sim
x^{(0)}-\rp{b}{3q},\eqf i.e. \eqi a\sim \left(\rp{\mu
P^2}{4\pi^2}\right)^{1/3}-\rp{\mu}{c^2}.\lb{newa}\eqf For HD
209458b the relativistic correction to the semimajor axis amounts
to $-1\times 10^{-8}$AU.

The modified mass function can be obtained by inserting \rfr{newa}
into the right-hand side of \rfr{vgu}. By using \eqi
a^3\sim\frac{\mu
P^2}{4\pi^2}-\frac{3}{c^2}\left(\frac{P}{2\pi}\right)^{4/3}\mu^{5/3},
\eqf it is straightforward to obtain \eqi \frac{K_M^3
P}{2G\pi}\sim\frac{m^3\sin^3
i}{(m+M)^2}\left[1-\frac{3}{c^2}\left(\frac{2\pi\mu}{P}\right)^{2/3}\right].\eqf
Thus one has $f=f^{(0)}+f^{(\rm GE)}$ with, to order
$\mathcal{O}(c^{-2})$,
\begin{eqnarray}f^{(0)}&=&\rp{m^3\sin^3
i}{(M+m)^2},\lb{clas}\\f^{(\rm GE)}&\sim &-\rp{3m^3\sin^3 i}
{c^2}\left(\rp{2\pi G}{P}\right)^{2/3} \left(\rp{1}{
M+m}\right)^{4/3}.
\end{eqnarray}
For HD 209458b $f^{(0)}=
2\times 10^{-10}$ M$_{\odot}$ and $f^{(\rm GE)}=
-1.5\times 10^{-16}$ M$_{\odot}$. The observational uncertainty in
$f$ can be evaluated as \eqi\sigma_{f^{\rm
(obs)}}\leq\left(\rp{K^3_M}{2\pi G }\right)\sigma_{P^{\rm (obs) }}
+\left(\rp{3PK^2_M}{2\pi G }\right)\sigma_{K_M}=
1\times 10^{-11}\ {\rm M}_{\odot}. \eqf We have assumed
$\sigma_{P^{\rm (obs)}}=0.01$ s (Wittenmyer et al. 2005), and
$\sigma_{K_M}=1.3$ m s$^{-1}$ (Wittenmyer et al. 2005).
This means that the relativistic corrections to the mass function
are too small to play a role in the parameters' determination. It
turns out that the major limiting factor is that due to the
uncertainty in
$K_M$
($1\times 10^{-11}$ M$_{\odot}$); the impact of $P^{(\rm obs)}$ is
smaller amounting to
$1\times 10^{-17}$M$_{\odot}$.

The error in the planet's mass $m$, which is measured from
\rfr{massf} and is mainly limited by $\sigma_{K_M}$ via $\sigma_f$
as $\sigma_m\sim \sigma_f/\sin^3 i$, also affects the measurement
of $a$. The uncertainty in $a$ is
\begin{equation}\sigma_a \leq  \rp{2}{3}\left[\rp{G(M+m)}{4\pi^2
P}\right]^{1/3}\sigma_{P^{\rm (obs) }}
+\rp{1}{3}\left[\rp{GP^2}{4\pi^2 (M+m)^2
}\right]^{1/3}\sigma_{m}=4\times 10^{-7}\ {\rm AU}.\end{equation}
We have used $\sigma_m=0.034$ m$_{\rm Jup}$ (Wittenmyer et al.
2005). The impact of $\sigma_m$
induces an error of $10^{-7}$ AU. The uncertainty in the measured
period yields an error of $10^{-9}$ AU. Thus, the relativistic
correction to $a$ is too small to be measured due to $\sigma_m$.
\section{The Systematic Error in the Keplerian Period}
The Keplerian period $P^{(0)}$ must be subtracted from the
measured value of $P^{\rm (obs)}$ in order to extract the
relativistic correction $P^{\rm (GE)}$, provided that the
uncertainty in $P^{(0)}$ is not larger than $P^{\rm (GE)}$
itself\footnote{Another possible source of systematic bias is
represented, in principle, by the classical correction $P^{(J_2)}$
due to the quadrupolar mass moment $J_2$ of the star (Iorio 2005).
However, by assuming the fiducial value $J_2\sim 10^{-6}$
(Miralda-Escud$\acute{\rm e}$ 2002) it turns out that
$P^{(J_2)}\sim -0.01$ s. }.

The precision with which $m$ can be determined is  important for
our purposes because $\sigma_m$ fixes, directly and indirectly via
$\sigma_a$ , the uncertainty in the knowledge of the Keplerian
period \eqi
\sigma_{P^{(0)}}\leq\pi\sqrt{\rp{a^3}{G(M+m)^3}}\sigma_m+3\pi\sqrt{\rp{a}{G(M+m)}}\sigma_a=4.145\
{\rm s }+4.157\ {\rm s} =8.302\ {\rm s }.\eqf Again, the error in
$m$ sets the limit of the measurability of the relativistic effect
which is 75 times smaller than $\sigma_{P^{(0)}}$ .
\section{Discussion and Conclusions}
In this paper we have shown that the accuracy with which it is
nowadays possible to determine the ephemerides of the transiting
extrasolar planet HD 209458b 'Osiris' would allow, in principle,
to measure the general relativistic correction to the orbital
period $P$. Indeed, the orbital period of Osiris is measured with
a $\sim 0.01$ s precision, while the relativistic prediction for
the correction $P^{\rm (GE)}$ to the Keplerian period $P^{\rm
(0)}$ amounts to $\sim 0.1$ s. It is likely that near-future more
accurate measurements of the orbital periods of some of the
exoplanets closest to their main stars will allow to extend this
possibility also to them. Indeed, one of the goals of the joint
CNES-ESA COnvection ROtation and planetary Transits (COROT)
mission
({\texttt{http://sci.esa.int/science-e/www/area/index.cfm?fareaid=39}}),
to be launched in 2006, is the discovery of new transiting giant
gaseous planets\footnote{Also transits due to smaller telluric
planets should be observed. An analogous NASA mission, called
KEPLER ({\texttt{http://www.kepler.arc.nasa.gov/}}), will be
orbited in 2008. Due to its higher altitude it will be able to
discover, among other things, also planets with larger semimajor
axes and periods up to two years.} close to their stars, i.e. with
periods ranging from a few days to three months. At present, the
relativistic component of the orbital periods of the so far
discovered exoplanets, other than HD 209458b, turns out to be too
small by one order of magnitude to be detected.

If it was really possible to extract such a tiny  slowing down of
the orbital motion of the planet it would be the first test of
general relativity, in its weak-field and slow-motion linearized
approximation, in a planetary system other than our Solar System.
Moreover, it would be the first measurement of such a consequence
of the Schwarzschild gravitoelectric component of the space-time
metric.

Unfortunately, in the case of HD 209458b the total error
$\sigma^{\rm (tot)}_{P^{\rm (GE) }}=\sigma_{P^{\rm
(obs)}}+\sigma_{P^{(0)}}$ is still larger than the investigated
effect. Indeed, it turns out that the uncertainties in the
planet's mass $m$ and semimajor axis $a$, due to the error in the
projected semiamplitude of the star's radial velocity $K_M$,
induce a mismodelling in the Keplerian period $\sigma_{P^{(0)}}$
of 8.302 s. One-two orders of magnitude improvements in
determining $K_M$ would be required.


\begin{thebibliography}{xxxxx}

\item []
Alonso, R. et al. 2004, ApJ,613, L153

\item []
Bodenheimer, P., Laughlin, G., \& Lin, D.N.C. 2003, ApJ, 592, 555

\item []
Charbonneau, D. et al. 2000, ApJ, 529, L45

\item []
Cody, A.M. \& Sasselov, D.D. 2002, ApJ, 569, 451

\item []
Estabrook, F. 1971, JPL IOM, Section 328, 1

\item []
Henry, G.W. et al. 2000, ApJ, 529, L41

\item []
Iorio, L. 2005, MNRAS, 359, 328


\item []
Konacki, M. et al. 2003, Nature, 421, 507

\item []
Mashhoon, B., Iorio, L. \& Lichtenegger, H.I.M. 2001,
{Phys. Lett. A}, {\rm 292}, 49

\item []
Mazeh, T. et al. 2000, ApJ, 532, L55

\item []
Miralda-Escud$\acute{\rm e}$, J. 2002, AJ, 564, 1019


\item []
Orosz, J.A., \& Hauschildt, P.H. 2000, A\&A, 364, 265


\item []
Rucinski, S. et al. 2003, Adv. Sp. Res. 31, 371

\item []
Schultz, A.B. et al. 2004, in {\it The Search For Other Worlds},
AIP Conf. Proc. 713, 161

\item []
Soffel, M.H. 1989, {\it Relativity in Astrometry, Celestial
Mechanics and Geodesy}, Springer

\item []
Wex, N. 1995,
{Class. Quantum Grav.} 12, 983

\item []
Will, C. M. 1993, {\it Theory and Experiment in Gravitational
Physics}, 2nd edition, Cambridge University Press, Cambridge

\item []
Wittenmyer, R.A. et al. 2005, ApJ 632, 1157




\end{thebibliography}
\end{document}